\documentclass{caosp}

\usepackage{graphicx}

\articleNo{123}
\pubyear{2014}
\volume{43}
\volnumber{1}
\firstpage{338}
\received{November 5, 2013}
\accepted{January 15, 2014}

\begin{document}

%
\hauthor{L. Szabados}

\title{Pulsating stars -- plethora of variables and observational tasks
}


%
\author{
        L.\,Szabados
       }

%
\institute{
Konkoly Observatory,
Research Centre for Astronomy and Earth Sciences,
Hungarian Academy of Sciences,
Budapest, Hungary\\
\email{szabados@konkoly.hu}
}

\date{November 5, 2013}

\maketitle

\begin{abstract}
Developments of this far-reaching research field are summarized from 
an observational point of view, mentioning important and interesting 
phenomena discovered recently by photometry of stellar oscillations of 
any kind. A special emphasis is laid on Cepheids and RR~Lyrae type variables.

\end{abstract}

%
\section{Introduction}
\label{intr}

Variable stars are astrophysical laboratories. Pulsating stars provide 
us with information on the internal structure of the stars and stellar 
evolution as testified by their position in the Hertzsprung-Russell (H-R) 
diagram. Hot and cool oscillating stars, and luminous and low 
luminosity pulsators are also found in various parts of the H-R diagram 
(Fig.~1). Several types of luminous pulsators are useful distance 
indicators via the period-luminosity ($P$-$L$) relationship.

\begin{figure}[!]
\centerline{\includegraphics[width=6.0cm,clip=]{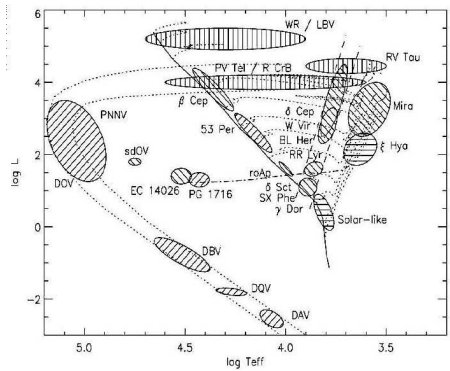}}
\caption{H-R diagram showing the location of various types of
pulsating variables (Jeffery, 2008a).}
\label{fig1}
\end{figure}

\begin{table}  
\footnotesize
\begin{center}  
\caption{Classification of pulsating variable stars.} 
\label{pulsvartypes}  
\begin{tabular}{l@{\hskip2mm}l@{\hskip2mm}l@{\hskip2mm}r@{\hskip2mm}c@{\hskip2mm}l} 
\hline\hline \\
Type &  Design. & Spectrum & Period & Amplitude & Remark$^\ast$\\
 & & & &  mag. & \\[0.5ex]
\hline \\
Cepheids    & DCEP   &  F-G Iab-II  & 1-135\,d   &  0.03-2  & \\
            & DCEPS  &  F5-F8 Iab-II &  $<$7\,d  &  $<$0.5  & 1OT \\
BL Boo      & ACEP   &  A-F         & 0.4-2\,d   &  0.4-1.0 & anomalous Cepheid\\
W Vir       & CWA    &  FIb         &    $>$8\,d &  0.3-1.2 & \\
BL Her      & CWB    &  FII         &    $<$8\,d &  $<$1.2  & \\
RV Tau      & RV     &  F-G         & 30-150\,d  &  up to 3 & \\
            & RVB    &  F-G         & 30-150\,d  &  up to 3 & variable mean brightness\\
RR Lyr      & RRA    &  A-F giant   & 0.3-1.2\,d &  0.4-2   & \\
            & RRC    &  A-F giant   & 0.2-0.5\,d & $<$0.8   & 1OT\\
$\delta$ Sct & DSCT  &  A0-F5\,III-V & 0.01-0.2\,d & 0.003-0.9 & R+NR\\
SX Phe      & SXPHE  &  A2-F5 subdw.& 0.04-0.08\,d &$<$0.7  &  Pop. II\\
$\gamma$ Dor & GDOR  &  A7-F7\,IV-V & 0.3-3\,d  & $<$0.1   &  NR, high-order g-mode \\
roAp        & ROAP   &  B8-F0\,Vp   & 5-20 min   & 0.01    &  NR p-modes\\
$\lambda$ Boo & LBOO &  A-F         & $<$0.1\,d  & $<$0.05 &  Pop.\,I, metal-poor\\
Maia        &        &  A           &            &         &  to be confirmed\\
V361 Hya    & RPHS,  &  sdB         & 80-600\,s  & 0.02-0.05 & NR, p-mode\\
            & EC14026 &             &            &         & \\
V1093 Her   & PG1716, & sdB         & 45-180 min & $<$0.02 &   g-mode\\
            & Betsy  &	            &            &         & \\
DW Lyn      &        &  subdwarf    &            & $<$0.05 &   V1093\,Her\,+\,V361\,Hya\\
GW Vir      & DOV,   &  HeII, CIV   & 300-5000\,s & $<$0.2 &   NR g-modes \\
            & PG1159 &              &            &         & \\   
ZZ Cet      & DAV    &  DAV         & 30-1500\,s & 0.001-0.2 & NR g-modes\\
DQV         & DQV    &  white dwarf & 7-18 min   & $<$0.05 & hot carbon atmosphere\\
V777 Her    & DBV    &  He lines    & 100-1000\,s &$<$0.2  & NR g-modes\\[0.5ex]
\hline\\
Solar-like  &        & F5-K1\,III-V & $<$hours  & $<$0.05  & many modes\\
\hspace*{5mm}oscill. &     &        &           &          & \\[0.5ex]
\hline\\
Mira        & M      &  M, C, S IIIe & 80-1000\,d & 2.5-11 & small bolometric ampl.\\
Small ampl. & SARV   &  K-M\,IIIe   & 10-2000\,d  & $<$1.0 & \\
\hspace*{5mm}red var.  &        &              &            &         & \\
Semi-regular & SR    &  late type I-III & 20-2300\,d & 0.04-2 & \\
            &  SRA   & M, C, S\,III & 35-1200\,d & $<$2.5  & R overtone\\
            &  SRB   & M, C, S\,III & 20-2300\,d & $<$2    & weak periodicity\\
	    &  SRC   & M, C, S\,I-II & 30-2000\,d & 1      & \\
            &  SRD   & F-K\,I-III   & 30-1100\,d & 0.1-4   & \\	    
Long-period &  L     & late type    &            &         & slow \\
\hspace*{5mm}irregular &   &        &            &         & \\
            &  LB    & K-M, C, S III &           &         & \\
            &  LC    & K-M I-III    &            &         & \\
Protoplan.  & PPN    &  F-G I       & 35-200\,d   &        & SG, IR excess\\
\hspace*{5mm}nebulae &  &           &             &        & \\[0.5ex]
\hline\hline
\end{tabular} 
\end{center}  
$^\ast$ R = radial; NR = non-radial; 1OT = first overtone; SG = supergiant.
Spectrum is given for maximum brightness for large amplitude variables.
\end{table}

\setcounter{table}{0}
\begin{table}  
\footnotesize
\begin{center}  
\caption{Classification of pulsating variable stars (continued).} 
\begin{tabular}{l@{\hskip2mm}l@{\hskip2mm}l@{\hskip2mm}r@{\hskip2mm}c@{\hskip2mm}l} 
\hline\hline \\
Type &  Design. & Spectrum & Period & Amplitude & Remark$^\ast$\\
 & & & &  mag. & \\[0.5ex]
\hline \\
53 Per      &        &  O9-B5       & 1-3\,d      &        & NR\\
$\beta$ Cep & BCEP   &  O6-B6\,III-V & 0.1-0.6    & 0.01-0.3 & R + NR\\
            & BCEPS  &  B2-B3\,IV-V & 0.02-0.04  & 0.015-0.025 & R + NR \\  
SPB         & SPB    &  B2-B9\,V    & 0.4-5\,d    & $<$0.5 & high radial order, \\
            &        &              &            &         & low degree g-modes\\
Be          & BE, LERI &Be          & 0.3-3\,d   &         & NR (or rotational?)\\
LBV         & LBV     & hot SG      & 30-50\,d   &         & NR?\\
$\alpha$ Cyg & ACYG    & Bep-Aep\,Ia  & 1-50\,d   & ~0.1    & NR, multiperiodic\\
BX Cir      &         & B           & ~0.1\,d    & ~0.1    & H-deficient \\						   
PV Tel      & PVTELI  & B-A\,Ip     & 5-30\,d    & ~0.1    & He SG, R strange mode\\
	    & PVTELII & O-B\,I      & 0.5-5\,d   &         & H-def. SG, NR g-mode \\     
	    & PVTELIII& F-G\,I      & 20-100 d	 &         & H-def. SG, R?\\[0.5ex]
\hline\hline
\end{tabular} 
\end{center}  
\end{table}	     

Table~1 is an overview of different types of pulsating variables. 
The underlying physical mechanism exciting stellar oscillations 
can be different for various pulsators. The General Catalogue of 
Variable Stars (GCVS, Samus et~al., 2009) lists 33 types and subtypes 
of pulsating variables, while the International Variable Star Index 
(VSX) at the AAVSO knows 53 different (sub)types. 

Another aspect of the classification is the ambiguity due to the simultaneous
presence of more than one type of variability. There are numerous pulsating
stars among eclipsing variables, as well as rotational variability can be 
superimposed on stellar oscillations. Pulsation can be excited in certain
cataclysmic variables, and erratic variability is typically present in 
oscillating pre-main sequence stars. From the point of view of astrophysics 
this is favourable but encumbers the analysis and interpretation of the 
observational data.

Time consuming photometry of pulsating variables is a realm of small 
telescopes. The temporal coverage (duration of the time series) is critical 
for studying multiperiodicity, changes in frequency content, modal 
amplitudes, etc. 

The accuracy of photometry is varying, it depends on the telescope
aperture, detector quality, astroclimate, etc. Millimagnitude accuracy
can be easily achieved with ground-based equipments, while the accuracy 
of photometry from space is up to micromagnitudes. Figure~2 shows an
excellent sample light curve of LR~UMa, a DSCT type pulsator obtained 
with a 1\,m telescope (Joshi et~al., 2000). (The abbreviated 
designation of various types is found in the 2nd column of 
Table~\ref{pulsvartypes}).

\begin{figure}
\centerline{\includegraphics[width=8.0cm,clip=]{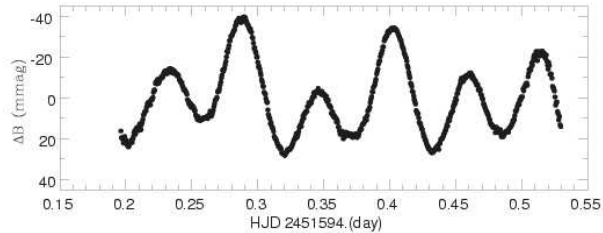}}
\caption{A very low-noise light curve of a short-period pulsator 
LR~UMa (Joshi et~al., 2000).}
\label{f2}
\end{figure}

Depending on the observer's experience and capabilities, one may choose
observational targets from a wide range of amplitudes, from microvariables
to large amplitude pulsators, while the range of periodicity embraces
the shortest values of seconds to the longest ones, several years.

In July 2013, there were 47811 variables catalogued in the GCVS, among them
8533 RR~Lyraes, 8098 Miras, 932 classical Cepheids, 762 $\delta$~Sct variables,
414 Type~II Cepheids, 209 $\beta$~Cep variables, 85 $\gamma$~Dor stars, and
80 white dwarf pulsators. A smaller number of variables known to belong to a 
certain type, however, does not necessarily mean that the given type of 
pulsating variables is less frequent, some kind of variability is not
easy to discover. Moreover, massive photometric surveys, e.g., ASAS
(Pojmanski, 2002), OGLE (Szyma\'nski, 2005), MACHO (Alcock et~al., 1999),
WASP (Pollacco et~al., 2006), and Pan-STARRS (Burgett \& Kaiser, 2009) 
resulted in revealing thousands of new variables not catalogued in the GCVS. 

\section{Remarkable behaviour of various pulsating variables}

In this section several interesting phenomena observed recently in
various types of pulsating stars are described. 

The $\gamma$~Doradus and $\delta$~Scuti stars are particularly useful for 
studying stellar structure and testing related theoretical models via
stellar oscillations (astero\-seismology). The GDOR stars pulsate in 
high-order g-modes with periods of order 1~day, driven by convective 
blocking at the base of their envelope convection zone. The DSCT stars 
pulsate in low-order g- and p-modes with periods of order 2~hours, 
driven by the $\kappa$ mechanism operating in the He\,II ionization zone. 
Theory predicts an overlap region in the H-R diagram between instability 
regions, where hybrid stars performing both DSCT and GDOR type pulsations 
should exist. Before the launch of Kepler spacecraft, only four such hybrid 
pulsators were known. From the period analysis of early Kepler data 
performed for more than 200 pulsators Grigahc\'ene et~al. (2010) found 
very rich frequency spectra and conclude that essentially all of the stars 
show frequencies in both the DSCT and the GDOR frequency range. 

The DSCT pulsation can be also coupled with solar-like oscillations excited
by envelope convection, as discovered in the case of HD\,187547 by Antoci 
et~al. (2011).

Even pre-main sequence stars located in the classical instability strip
can pulsate resulting in superposition of erratic and periodic components
of stellar variability. As an example, the study of pre-main sequence
pulsators in the young clusters IC 4996 and NGC 6530 is mentioned 
(Zwintz \& Weiss, 2006).

A particularly interesting DSCT pulsator is WASP-33 being the host star of
an extrasolar planet with an orbital period of 1.21987 days
(Herrero et~al. 2011). None of the observed pulsation frequencies, nor their 
low order linear combinations are in close resonance with the orbital frequency
(Kov\'acs et~al. 2013).	Stability of the oscillation frequencies and 
amplitudes is well worth studying on a longer time scale.

Discovery of DSCT and GDOR pulsation was even discovered in Ap stars 
from Kepler data (Balona et al., 2011). Such periodic pulsation of Ap stars
was unknown before. On the contrary, rapid oscillations in Ap stars have
been known for decades but their cause have not been clarified yet.
The presence of pulsation in Ap stars has to do with the internal stellar
magnetic field. About 40 roAp stars are known, i.e., not all Ap stars are
rapid pulsators. Not oscillating Ap stars (noAp stars) occupy a similar 
part of the H-R diagram as the roAp stars.

There is a wide variety of pulsating stars among hotter, B type stars.
Here again, one can find hybrid pulsators. The `classical' $\beta$~Cephei
pulsation can be coupled with either SPB or Be type variability. In the
case of pulsation of Be stars, long-term coherent photospheric oscillations
may be present accompanied with quasi-periods of circumstellar origin
due to a mass ejection episode in the rapidly rotating hot star, as 
in the of HD~50064 (Aerts et~al., 2010b). Another kind of hybrid pulsation
is the simultaneous presence of BCEP and SPB type oscillations in the same 
stars: in the case of $\gamma$~Peg, a large number of high-order g-modes, 
low-order p-modes and mixed modes have been detected by Handler et~al. (2009).

Slow pulsation of B stars, i.e., the SPB pulsation is not a rare phenomenon, 
though its discovery is not easy. The precise and homogeneous Hipparcos 
photometry was instrumental in revealing a large number of SPB pulsators, 
and more recently McNamara et~al. (2012) found dozens of new SPB variables 
in the Kepler field. SPB pulsation may be present among pre-main sequence 
variables, as well (Gruber et~al. 2012). 

Even a supergiant B star may show SPB type variability as revealed 
in HD~163899 (Saio et al., 2006) from MOST photometry. This is
surprizing and needs an appropriate pulsation model. Another new
type of variability among B stars was discovered by Mowlavi et~al. (2013)
who found a number of new variable stars between the red edge of 
SPB instability region and the blue edge of DSCT stars, where no pulsation 
is predicted to occur based on the existing stellar models.

Pulsation is present in the most luminous stars, see the light curve 
of the luminous blue variable (LBV) AG~Car in Aerts et~al. (2010a).

The least luminous stars can also pulsate in a variety of locations
of the white dwarf and subdwarf regions of the H-R diagram: the ZZ~Cet 
type pulsation of white dwarfs has been known from 1968, more recently 
discovered types are GW~Vir (1979), V777~Her (1982), while the types of 
subdwarf pulsators are V361~Hya (1997) and V1093~Her (2002). Moreover, 
DW~Lyn type (2002) is a hybrid of V361~Hya and V1093~Her type pulsations.

There are less than 15 extreme He-stars known to pulsate in our Galaxy
(PV~Tel, BX~Cir types, Jeffery, 2008b). Study of such variables are 
informative on late stages of stellar evolution.

RCRB stars falling within the classical instability strip also pulsate,
including the archetype R~CrB itself (Rao \& Lambert, 1997). Crause et~al. 
(2007) put forward convincing evidence that the decline events (i.e.,
the mass-loss episodes) occurring in RCRB variables are synchronized to 
the atmospheric oscillations.

The peculiar variable star FG~Sge (a post-AGB central star of a planetary
nebula) also showed periodic pulsation while the temperature of its 
false photosphere was appropriate during the rapid crossing of the
instability strip and before becoming a cool RCRB variable (Jurcsik 
\& Montesinos, 1999). This sequence of events was a stellar evolutionary 
episode on a human time scale.

There are numerous pulsating variables among red stars, as well.
Some of the Mira variables are also famous of undergoing rapid
evolutionary episode of He-shell flash, including T~UMi (Szatm\'ary 
et~al., 2003) and R~Cen (Hawkins et~al., 2001). The pulsation
period of these stars is noticeably decreasing from one cycle to
the other accompanied by a secular decrease of the pulsation
amplitude. Hawkins et~al. (2001) also revealed a correlation 
between the instantaneous period and semi-amplitude of the pulsation
of R~Cen. Other cases of secular evolution observed in the pulsation of 
Mira variables are listed by Templeton et~al. (2005).

Pulsating stars in the post-AGB phase of stellar evolution, e.g., RV~Tau
type variables also show interesting phenomena in their photometric 
behaviour (Kiss et~al., 2007).

\section{Importance of binarity among pulsating variables}

Binarity is important in variable star research: 
both eclipsing and cataclysmic phenomena are caused by the
presence of a companion star. In the study of pulsating variables,
binarity provides an additional aspect to be taken into account
in interpreting the observed variability.

On the one hand, a luminous companion can decrease the observable
photometric amplitude of the pulsating component, contributing to 
the wide range of amplitude of Cepheids observed at a given pulsation
period. On the other hand, a close (and not necessarily luminous) 
companion can even trigger stellar oscillations in the other star
of the binary system. This is the case of the `heartbeat' variable, 
KOI-54 (HD\,187091) discovered by Welsh et al. (2011) in
the Kepler field (Fig.~3). There is a whole class of eccentric 
binaries in which pulsations are excited tidally (Thompson et~al., 
2012).

\begin{figure}
\centerline{\includegraphics[width=11cm,clip=]{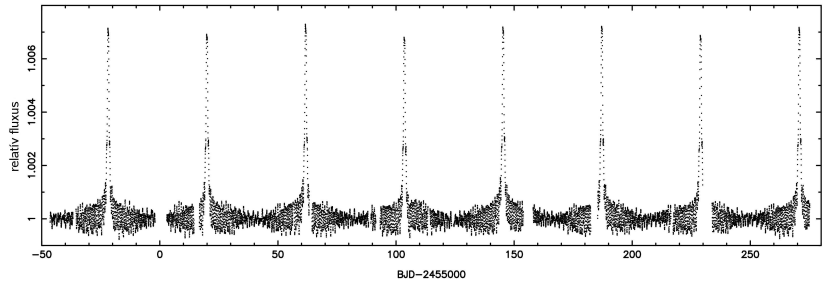}}
\caption{Light curve of the heartbeat variable KOI-54 (HD\,187091)
(Welsh et~al., 2011).}
\label{f3}
\end{figure}

Another kind of externally triggered pulsation was observed in the
symbiotic nova RR~Tel preceding its eruption in 1948 (Robinson, 1975).

Long-period variations in the mean brightness of RV~Tauri stars
(RVB subtype) is also caused by the binarity of these pulsators.

Pulsating variables in binary systems can show apparent period
changes owing to the light-time effect caused by the orbital
motion. Such effect was revealed in the $O-C$ diagram of different
types of pulsators, e.g., the archetype BCEP star $\beta$~Cep
(Pigulski \& Boratyn, 1992), the DCEP variable AW~Per (Vink\'o, 1993),
the DSCT star SZ~Lyn (Derekas et al., 2003), the SXPHE star
CY~Aqr (Sterken et al., 2011), and the SPB variable HD\,25558 
(S\'odor et al., in preparation). The light-time effect is
instrumental in determining orbital elements of the given binary
system. In the previous list, HD\,25558 is a unique system whose
both components are SPB variables.

Binarity is an important aspect for the calibration of the 
$P$-$L$ relationship. On the one hand, the photometric
contribution of the companion star has to be removed when
determining the luminosity of the pulsator involved in the
calibration procedure. On the other hand, pulsating stars
in binaries with known orbital elements are useful calibrators
on their own right.

\section{Strange behaviour of classical Cepheids}

Astronomy textbooks usually refer to Cepheids as regular 
radial pulsators with strongly repetitive light curves.
However, recently it turned out that classical Cepheids
are not perfect astrophysical clocks.
V1154~Cyg, a first overtone pulsator, the only Cepheid in the 
Kepler field shows cycle-to-cycle variations in both the shape
of its light curve and pulsation period (Derekas et~al., 2012).
This behaviour needs an astrophysical explanation. Though the 
instantaneous pulsation period flickers (Fig.~4), the average 
period remains stable on the time scale of several decades.	

\begin{figure}
\centerline{\includegraphics[width=8.5cm,clip=]{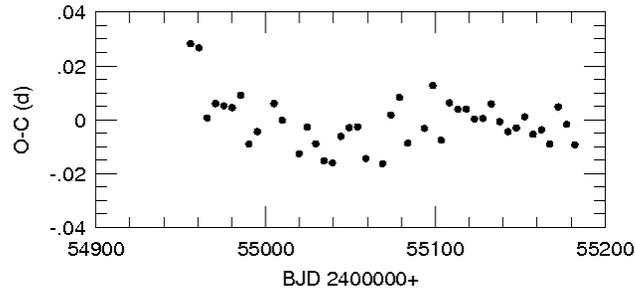}}
\caption{$O-C$ diagram of the Cepheid V1154 Cygni based on the Kepler
photometry (Derekas et~al., 2012).}
\label{f4}
\end{figure}

Stellar evolution has its impact on the pulsation period of
Cepheids, as well. Rapidly evolving long period Cepheids sometimes
show spectacular changes in their pulsation period, and erratic 
fluctuations superimposed on the secular period variation (Fig.~5).

\begin{figure}[thp]
\centerline{\includegraphics[width=5.5cm,clip=]{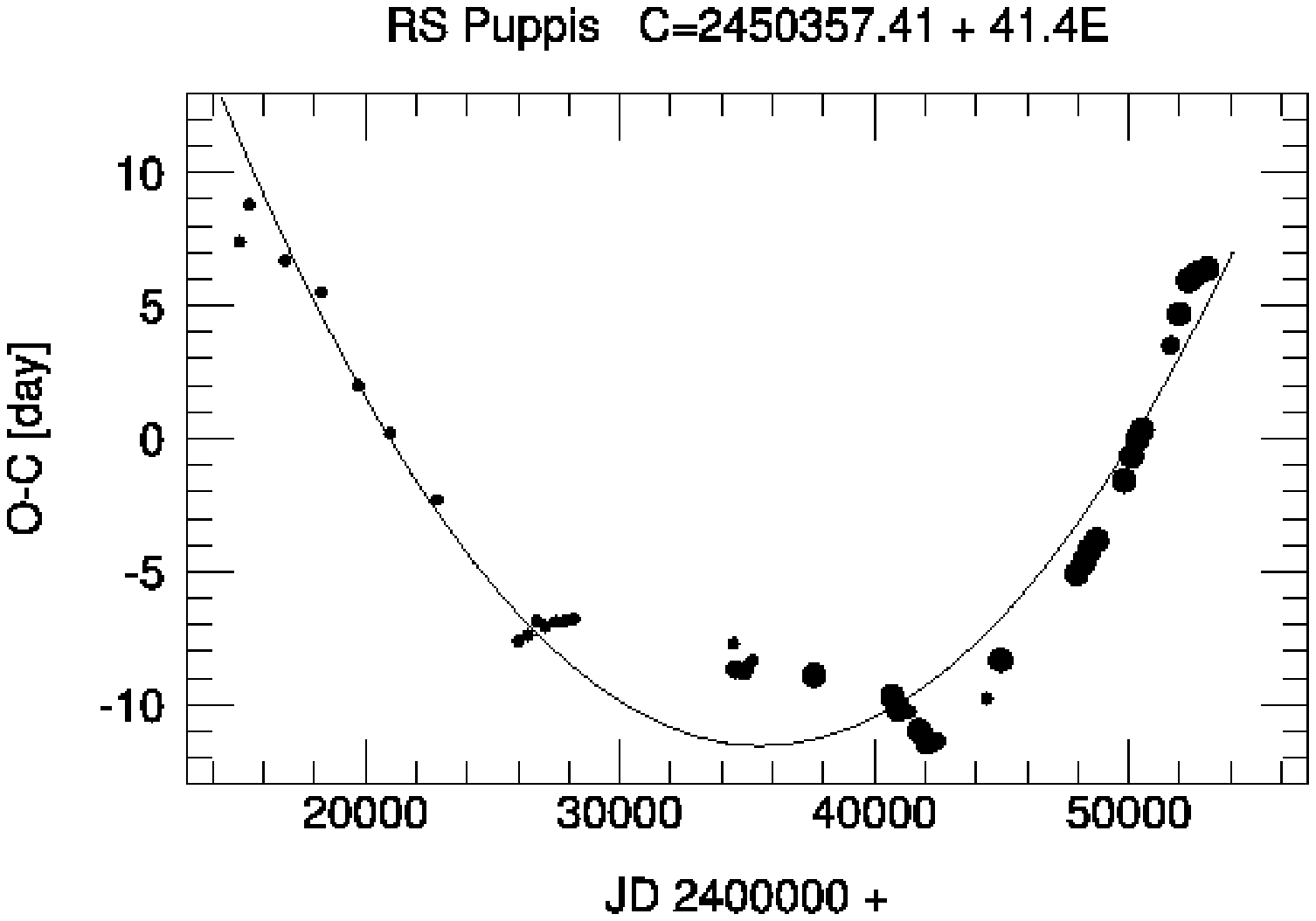} 
            \hspace{0.5cm}
            \includegraphics[width=5.5cm,clip=]{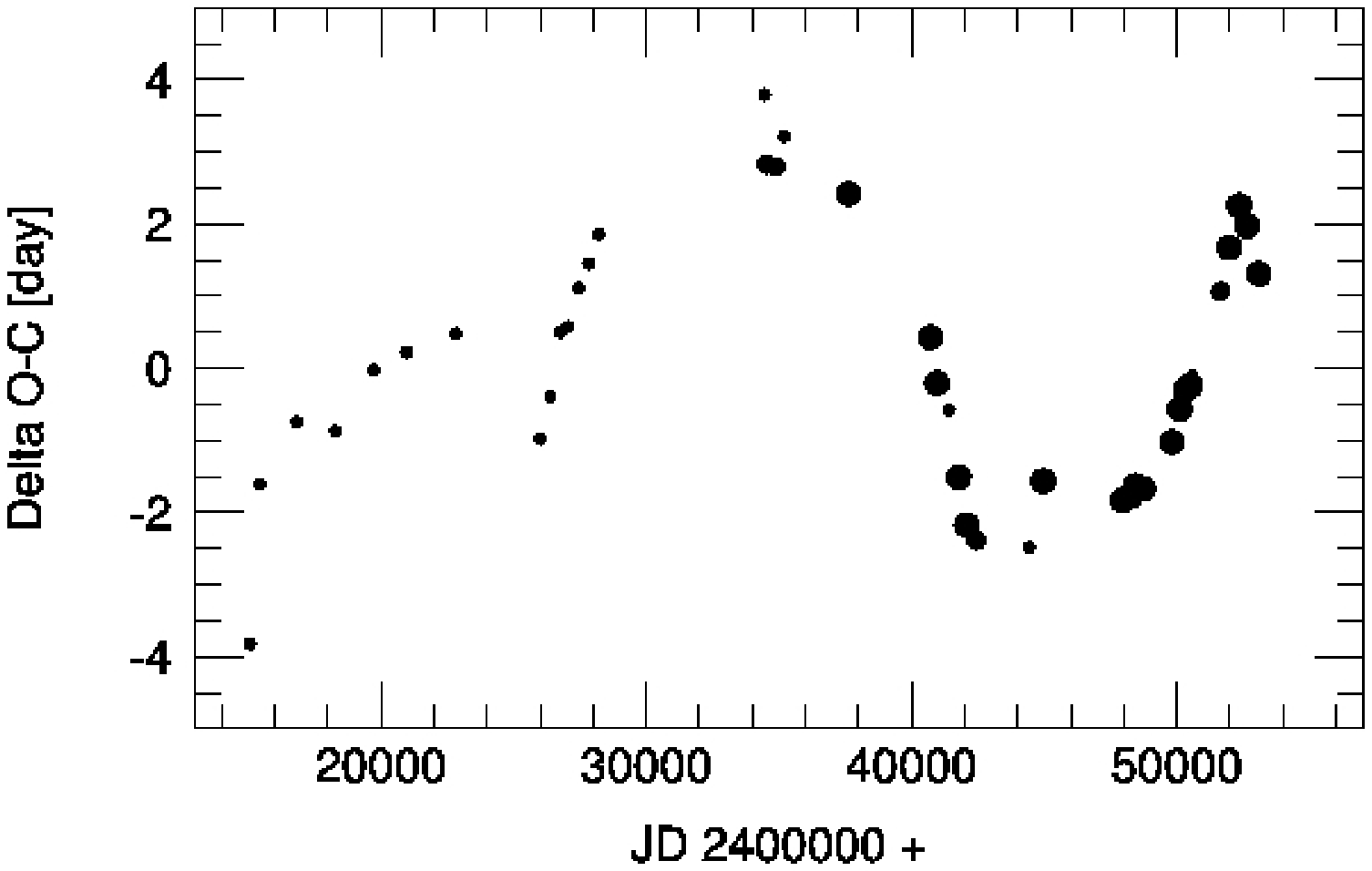} }
\caption{$O-C$ diagram of the long-period Cepheid RS~Puppis (left) 
and the residuals after subtracting the fitted parabola (right)
(Szabados, unpubl.).}
\label{f5}
\end{figure}

Subtle period changes can be pointed out from long series of
high-quality photometric observations. The definite phase jump
in the pulsation of the peculiar Cepheid Polaris is especially
noteworthy (Turner et~al., 2005). This jump may be a result of
a proximity effect in the binary system. The origin of the
secular variation in the pulsation amplitude of Polaris is, 
however, a mystery.

An additional periodicity is frequently present in 
Cepheids: hundreds of double-mode Cepheids are known in both
Magellanic Clouds. In these galaxies there exist Cepheids
pulsating simultaneously in three radial modes. Slight excitation
of nonradial modes was also found in 9\% of the firt overtone 
Cepheids in the Large Magellanic Cloud, and signs of the Blazhko 
effect (a typical phenomenon of RR~Lyrae type pulsators) have also 
been revealed among Magellanic double-mode Cepheids (Moskalik, 2013).
In our Galaxy, there is only one known Cepheid showing the Blazhko 
effect: V473~Lyr (Moln\'ar et al., 2013).

\section{RR Lyrae variables}

In addition to their large amplitude periodic pulsation, RR~Lyr type
variables also show various other effects worthy of studying in
detail. The most frequently occurring phenomenon is the Blazhko effect,
a slow, cyclic (not periodic) modulation of the light curve (both
amplitude and phase) observed in both RRA and RRC variables (Fig.~6). 
In spite of the fact that Blazhko effect occurs in about 50\% of the 
field RRab stars, its origin is a century-long enigma. Although several 
models have been elaborated (magnetic oblique rotator, nonradial 
resonant rotator, interaction of shock waves, cycles in the convection), 
none of them can be accepted as a real explanation.

An up-to-date list of RR~Lyrae variables in the Galactic field known 
to exhibit Blazhko effect has been compiled by Skarka (2013).
This catalog of Blazhko modulated RR~Lyr stars contains 242 variables 
including 8 stars with more than one (incommensurable) modulation period, 
and 4 stars whose modulation period strongly varies.

\begin{figure}
\centerline{\includegraphics[width=12cm,clip=]{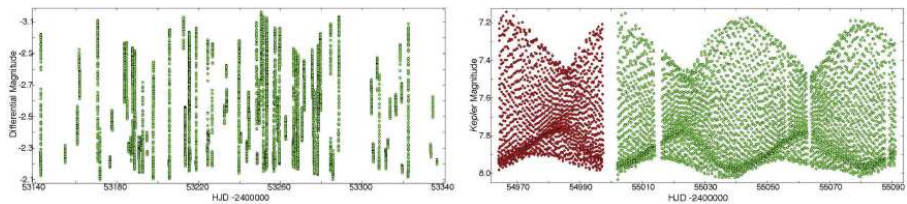}}
\caption{Light curve of RR~Lyr from ground-based (left) and 
space-based (Kepler) data (right)(adapted from Kolenberg, 2011).}
\label{f6}
\end{figure}

Double-mode pulsation and nonradial modes are also present in some
RR type variables (Moskalik, 2013).

Based on Kepler data new dynamical phenomena have been discovered:
period doubling (in RR~Lyr by Moln\'ar et~al., 2012), triple-mode 
pulsation (in V445~Lyr by Guggenberger et~al., 2012; in RR~Lyr by 
Moln\'ar et~al. 2012), and high-order resonances (in RR~Lyr by 
Moln\'ar et~al. 2012). It is promising that new models involving 
interactions between radial and nonradial modes of oscillation
as well as coupling between the fundamental mode, first overtone 
and a high-order (9th) radial mode can lead us to the correct 
explanation of the Blazhko effect present in RR~Lyrae variables.

\section{Plethora of optical telescopes}

\begin{table}  
\small
\begin{center}  
\caption{Space telescopes used for or dedicated to optical photometry.} 
\label{spacephotom}  
\begin{tabular}{l@{\hskip2mm}l@{\hskip2mm}c@{\hskip2mm}l} 
\hline\hline  
Mission &  Duration &  Aperture (cm) & Remark\\
\hline  
IUE      & 1978-1996 &  45 &  FES, no calibration\\
Hipparcos& 1989-1993 &  29 &  Hp wide-band magnitude\\
HST      & 1990-     & 240  &  FGS\\
WIRE     & 1999-2011 &   5.2&  star tracker\\
INTEGRAL & 2002-     &   5  &  OMC, Johnson $V$\\
Coriolis & 2003-     &   1.3&  SMEI  instrument\\
MOST     & 2003-     &  15  &  limited field (CVZ)\\
CoRoT    & 2006-2012 &  27  &  very limited field\\
Kepler   & 2009-2013 &  95  &  very limited field\\
BRITE    & 2013      &   3  &  blue \& red bands\\
\hline\hline  
\end{tabular} 
\end{center}  
\end{table}

Figure~1 in the paper by Mountain \& Gillett (1998) (not repeated here)
shows the temporal increase of the cumulative mirror area of optical 
astronomical telescopes. Owing to the enormous progress in engineering, 
there exist giant telescopes in service of astronomy, yet small aperture 
telescopes contribute overwhelmingly to the recent steep increase. 
These small telescopes (up to 1.5\,m diameter) are ideal instruments 
for carrying out photometric observations of variable stars.

In addition to ground-based equipments, there exist photometric space 
telescopes or other space telescopes also used for stellar photometry
(see Table~2). Most of them have a small aperture, and it is favourable 
that their photometric data are accessible in most cases.

\section{Plethora of new variables}

Although the catalogued variable stars offer a wide choice for
photometric observers, there are many recently discovered pulsating
variable stars whose variability was revealed in massive photometric 
surveys, e.g., ASAS, OGLE, \linebreak MACHO, VVV, and these variables 
have not been included in the GCVS yet.

Photometric data bases of major ground-based sky surveys such as
Catalina Sky Survey (Drake et~al., 2009), Pan-STARRS, Sloan Digital 
Sky Survey\linebreak (Abazian et~al., 2003), LSST (LSST Science 
Collaborations, 2009) are or will be ample sources of new targets 
with variable brightness (including pulsating variables) for thorough 
photometry with small or medium aperture telescopes.

In the coming years, discovery of a tremendous number of new
variable stars is envisaged. Gaia, the ESA's astrometric space probe 
will collect photometric data of a billion stars from 2014 on.
As a result, discovery of 18 million new variable stars is expected
from its data base. The estimated number of pulsating variable stars 
to be observed by Gaia is as follows (Eyer \& Cuypers, 2000):
2000-8000 Cepheids (9000 according to Windmark et~al., 2011),
70000 RR, 60000 DSCT, 140000-170000 Miras, 100000 SR stars, 3000 BCEP
variables, 15000 SPB pulsators, etc.

\section{Variety of observational tasks and their outcome}

It is not necessary to be involved in a long-term observational
project. Even a single light curve provides useful pieces of
information: for Cepheids and RR~Lyrae stars, the atmospheric
metallicity can be determined from the shape of the light curve
via Fourier decomposition (Klagyivik et~al., 2013), and the value
of the pulsation period can be updated with the help of the $O-C$ 
method, if prior photometric data are available.

From a data set obtained during one season one can determine the 
type of variability for newly revealed variable stars. Even 
discovery of a new type is possible, e.g., brown dwarf pulsation is 
predicted by theory (Palla \& Baraffe, 2005) but it has been 
unobserved yet.

From longer data sets, i.e. detailed photometric study of individual 
variables one can point out additional periodicities, perform a mode 
identification, discover slightly excited non-radial (or radial) modes.
Existence of triple-mode radial pulsators has been an unexpected recent
discovery (see e.g., Wils et al., 2008 and Moskalik, 2013).

Observations of pulsating variables in binary systems can be
especially fruitful because of interactions of binarity and pulsation
phenomena.

The astrophysical interpretation of the photometric data includes
the determination of physical properties of the given star(s) from the 
analysis of the light variations: evolutionary state, internal structure,
metallicity, rotation, presence of companion(s), etc.

It may happen that photometric data are insufficient for a reliable
analysis, yet the light variations indicate that the given pulsator
deserves an in-depth (spectroscopic) study with a larger telescope.
Cooperation between several telescopes/observatories is beneficial
in any case.

For further information about pulsating variable stars, the following
books are recommended: Aerts et~al. (2010a), Balona (2010), Percy (2007), 
and Su\'arez et~al. (2013) in which a lot more interesting phenomena 
have been discussed.

\acknowledgements
The organizers of the conference are thanked for dedicating an invited
review to this topic. The author acknowledges the anonymous referee's and 
M\'aria Kun's constructive comments on the manuscript. Financial support 
by the Hungarian OTKA grant K83790 and the ESTEC Contract No.~4000106398/12/NL/KML 
is gratefully acknowledged.


\end{document}